\documentclass[12pt,letterpaper]{article} 

\usepackage[english]{babel} 
\usepackage[utf8x]{inputenc} 

\usepackage{lipsum} 
\usepackage{setspace} 
\usepackage{microtype} 

\usepackage[margin=1.5in]{geometry} 
\usepackage{changepage} 

\usepackage{amsmath} 
\usepackage{amssymb} 
\usepackage{bm} 
\usepackage[retainorgcmds]{IEEEtrantools} 

\usepackage{graphicx} 
\usepackage[font=normalsize]{caption} 
\usepackage{subcaption} 

\usepackage{booktabs} 
\usepackage{tabularx} 
\usepackage{longtable} 
\usepackage{adjustbox} 
\usepackage{tabularray} 

\usepackage{listings} 
\usepackage{verbatim} 

\usepackage[colorlinks=true]{hyperref} 
\hypersetup{ 
  colorlinks,
  citecolor=blue,
  linkcolor=red,
  urlcolor=cyan
  }

\usepackage[colorinlistoftodos]{todonotes} 
\usepackage{soul} 
\sethlcolor{yellow} 

\usepackage{enumitem} 
\usepackage{pifont} 

\usepackage{xcolor} 
\usepackage{float} 
\usepackage{ragged2e} 
\usepackage{lineno} 
\usepackage{multicol} 
\usepackage{pdflscape} 
\usepackage{lscape} 
\usepackage[super]{nth} 

\usepackage{natbib} 

\usepackage{authblk}

\usepackage{fancyhdr} 
\fancyhfoffset[L]{0cm} 
\fancyhfoffset[R]{0cm} 
\fancypagestyle{abstractstyle}{
    \fancyhf{} 
    \fancyfoot[C]{\textcolor{gray}{\thepage}}

}

\usepackage{graphicx}%
\usepackage{multirow}%
\usepackage{amsmath,amssymb,amsfonts}%
\usepackage{amsthm}%
\usepackage{mathrsfs}%
\usepackage[title]{appendix}%
\usepackage{xcolor}%
\usepackage{textcomp}%
\usepackage{manyfoot}%
\usepackage{booktabs}%
\usepackage{algorithm}%
\usepackage{algorithmicx}%
\usepackage{algpseudocode}%
\usepackage{listings}%
\usepackage[table]{xcolor}
\usepackage{subcaption}
\usepackage{tabularx}

\usepackage{array}

\begin{document}


\title{
\vspace{-2cm}
\textbf{A study on healthcare expenditure in Italian regions via Symbolic Regression} 
}

\newcommand{\thetitle}{A study on healthcare expenditure in Italian regions via Symbolic Regression}

\author{Rosseti Cristina$^{a,b,\dag,*}$} 
\author{Nicola Caravaggio$^{c}$}
\author{Giovanni Lamura$^{d}$} 
\author{Giuliano Resce$^{c}$} 

\affil{\footnotesize
  $^{a}$\textit{Department of Information Engineering, Marche Polytechnic University, Ancona -- Italy}\\ 
  $^{b}$\textit{Department of Computer Science and Automation, Polytechnic of Turin, Turin -- Italy}\\ 
  $^{c}$\textit{Department of Economics, University of Molise, Campobasso -- Italy}\\ 
  $^{d}$\textit{Scientific Institute for Hospitalization and Care (INRCA), Ancona -- Italy} 
  \vspace{0.3cm}
  \begin{minipage}{11cm}~\\
  \centering
  \footnotesize
  $^{\dag}$\tt\href{mailto:c.rossetti@pm.univpm.it}{c.rossetti@pm.univpm.it} 
  \vspace{-1cm} 
  \end{minipage}
  }

\date{\small\today}

\renewcommand\Authands{ and }

\maketitle


\begin{adjustwidth}{-0.25in}{-0.25in} 
\begin{abstract}

\thispagestyle{abstractstyle}

\noindent\rule[0.5ex]{\linewidth}{0.5pt}

The study of the factors driving the dynamics of healthcare spending is of paramount importance to guide policymakers in the allocation of resources and to measure the effectiveness of the healthcare system under consideration. The identification of these drivers can be supported by the use of machine learning techniques, which enable the discovery of hidden patterns within vast amounts of data. In contrast to black-box methods, Symbolic Regression (SR) is an approach that allows for the identification of analytical models that explicitly capture the functional relationships within the data, thus enhancing interpretability. In this paper, we present the use of SR for identifying the drivers of healthcare expenditure in Italian regions. Given the dynamic and complex nature of this phenomenon, we generated several models based on distinct temporal windows, enabling us to analyze the drivers across different time horizons. In addition to identifying the main drivers based on variable frequency in the generated models, we also conducted a study on recurring substructures.  The results show that SR was able to generate models with a good level of predictive accuracy for private healthcare expenditure, enabling a reliable analysis of its driving factors, but failed to do so for public healthcare expenditure.

\noindent\rule[0.5ex]{\linewidth}{0.5pt}

\end{abstract}
\end{adjustwidth}

\begin{quote}

	
 
 
\end{quote}

\pagebreak


\onehalfspacing 

\section{Introduction}\label{sec:introduction}

Healthcare expenditure is a complex, multi-dimensional phenomenon influenced by a wide range of economic, demographic, and institutional factors. The empirical literature has identified several key determinants, including income levels \citep{roberts2000spurious}, the share of older population \citep{gerdtham1992econometric}, technological advancements \citep{white2007health}, and labor market conditions \citep{crivelli2006federalism}. Despite extensive research, the relative importance and interaction of these drivers remain difficult to disentangle, partly due to nonlinear relationships and structural heterogeneity across countries and over time.\\
Understanding the determinants of healthcare expenditure is of primary importance for both policymakers and researchers. On the one hand, healthcare spending represents a substantial and growing share of public budgets, raising concerns about fiscal sustainability. On the other hand, it reflects underlying population needs and system performance, making it a key indicator for evaluating health policy effectiveness. Accurately identifying its drivers is therefore crucial for designing efficient and equitable healthcare systems.\\
In this context, recent advances in machine learning (ML) methodologies offer new opportunities to model complex relationships without imposing strong \textit{a priori} assumptions. However, despite recent efforts to identify the drivers of healthcare spending using ML techniques \citep{irvin2020incorporating,vimont2022machine,caravaggio2026determinants}, many studies still rely on black-box algorithms, raising concerns about interpretability and limiting their usefulness for policy-oriented analysis. This paper addresses this gap by employing Symbolic Regression (SR), a supervised ML technique whose goal is to retrieve an analytical expression that best fits the input data. Since the relationships between the dependent variable and independent variables are made explicit, SR can retrieve interpretable models \citep{angelis2023artificial}. This has made SR of increasing interest to the research and scientific community \citep{dong2025recent}, including application domains such as healthcare and social sciences \citep{fong2024symbolic, truscott2014explaining}. As such, SR represents an interpretable ML approach that combines the flexibility of data-driven modeling with the transparency of analytical expressions, with the aim of uncovering meaningful and interpretable relationships between healthcare expenditure and its underlying determinants. 
In this regard, it can be useful to remember that interpretability can be broadly defined as the ability to easily explain an ML model or solution to humans \citep{doshi2017towards} or, more specifically, as the possibility to extract \textit{``relevant knowledge from a model about relationships contained in the data or learned by the model" }\citep{murdoch2019definitions}. 
In many high-stakes contexts, such as healthcare or criminal justice, where models are used to inform decisions, the use of interpretable models is of paramount importance
\citep{rudin2019stop, markus2021role, amarasinghe2023explainable}. This makes interpretable modeling particularly relevant for healthcare expenditure analysis. Indeed, identifying the relevant relationships and drivers of this phenomenon can help guide policymakers in understanding the factors that drive spending over time, allocating resources accordingly, and adopting specific strategies. \\
Motivated by these considerations, this study adopts SR in the context of healthcare expenditure in Italian regions, with the aim of retrieving interpretable models that might help human experts to better inform policy-making through the detection of explicit analytical relationships between different drivers. The expressions are intended to support expert exploration of observational patterns rather than to estimate the effects of changing those indicators. \\
Since our analysis covers a large temporal horizon, particular care must be taken to reduce the risk of data leakage on panel data, in order to avoid overly optimistic results by mixing past and future observations \citep{cerqua2024mis}, thus weakening the reliability of the investigated relationships. Moreover, healthcare expenditure is a complex phenomenon, whose dynamics may depend on historical time periods. As such, its drivers can change over the years, requiring special attention to the training of a single model. For these reasons, we carefully split data for training and testing and created distinct temporal windows over different year intervals. \\
In light of the above, the contributions of this paper can be summarized as follows:
\begin{itemize}
    \item using SR to retrieve explicit analytical expressions relating  healthcare expenditure to  socio-economic indicators. Rather than focusing only on predictive performance, we examine recurrent predictors and symbolic combinations across candidate models;
    \item explicitly addressing the problem of data leakage in temporally structured socio-economic data. This paper adopts a multiple splitting strategy in which SR is performed separately on different temporal windows, in order to avoid mixing temporal information between training and test sets and to carefully address changes in the relevance of drivers across different time horizons;
    \item evaluating the temporal transferability of the recovered expressions through out-of-time evaluation;
    \item analysing the stability of SR results across repeated runs, thus allowing us to detect the most common retrieved features and functional substructures. 
\end{itemize}
The paper is organized as follows. Section \ref{sec:literaturereview} presents a review of the literature on the determinants of healthcare expenditure, followed by a discussion of SR applications and issues related to data leakage in ML approaches. Section \ref{sec:methodology} describes the methodology adopted. Section \ref{sec:results} reports the results of the analysis, which are then discussed in Section \ref{sec:discussion}. Finally, Section \ref{sec:conclusion} formulates some concluding remarks.

\section{Evidence from the literature}
\label{sec:literaturereview}

The empirical literature on healthcare expenditure identifies income, demographic structure, technological progress, and institutional factors as the main determinants of spending dynamics. Early cross-country analyses, notably \cite{kleiman1974determinants} and the seminal contribution of \cite{newhouse1977medical}, documented a strong positive relationship between per capita GDP and healthcare expenditure, frequently reporting income elasticities greater than one and thus characterizing healthcare as a luxury good. This result was initially confirmed by subsequent studies that extended the baseline model by including demographic and system-level controls, such as \cite{leu1986public} and \cite{gerdtham1992econometric}. However, later econometric critiques highlighted methodological limitations related to non-stationarity and spurious correlations, questioning the robustness of early findings \citep{hansen1996determinants,blomqvist1997health,okunade2001unit}.
With the adoption of panel data techniques and improved data coverage, more recent studies generally report income elasticities at or below unity, particularly for OECD and high-income countries, suggesting that healthcare should currently be considered a necessity rather than a luxury good \citep{gerdtham1998determinants,martin2011review,baltagi2010health}. Although some contributions still find elasticities above one in specific contexts or periods \citep{roberts2000spurious,clemente2004international}, global evidence increasingly points to a declining role of income as the dominant driver of healthcare expenditure \citep{baltagi2017health}.
Population aging has also been widely analyzed as a determinant of healthcare spending. Several studies document a positive association between older age structures and higher healthcare expenditure \citep{gerdtham1992econometric,dimatteo1998evidence,karatzas2000determination}, while others find a more limited impact at the macro level \citep{getzen1992population,barros1998black}. A substantial strand of the literature shifts the focus from age to proximity to death, giving rise to the “red herring” hypothesis, according to which healthcare costs are driven primarily by expenditures concentrated in the final years of life, rather than by aging \textit{per se} \citep{zweifel1999ageing,felder2000health,shang2008does}. While this hypothesis is supported by extensive micro-level evidence, conflicting results persist, especially in country-specific studies where age retains explanatory power \citep{atella2014effect,karlsson2014testing}.
Technological progress is commonly identified as a major driver of healthcare expenditure growth, often measured indirectly as a residual component once income and demographic effects are accounted for \citep{newhouse1992medical,white2007health}. Empirical contributions using proxies such as medical equipment, research and development expenditure, and time trends consistently find a positive and significant impact of technological change on healthcare spending \citep{okunade2002technology,murthy2016determinants}. Finally, a broad set of additional determinants has been explored, including health system organization and decentralization \citep{mosca2007decentralization,lopez2007multilevel}, labor market conditions \citep{crivelli2006federalism}, and supply-side characteristics \citep{koenig2003drivers}. Overall, the literature underscores the multifactorial nature of healthcare expenditure and the absence of a single dominant explanatory factor.
The Italian literature largely confirms the relevance of income, demographic factors, and institutional settings in explaining healthcare expenditure. Early regional-level evidence by \cite{giannoni2002regional} shows that decentralization did not increase healthcare spending, while real GDP emerged as the main determinant, albeit with a low elasticity (0.33), consistent with healthcare being a normal good in Italy. Population aging played a secondary role, and supply-side variables suggested the presence of economies of scale alongside labor-intensive cost pressures. These findings are broadly confirmed by \cite{lagravinese2014corruption}, who also highlight the importance of institutional quality.
Focusing on demographic dynamics, \cite{lopreite2017effects} analyze Italian healthcare expenditure between 1990 and 2013 using a Bayesian framework, showing that aging exerts a stronger influence than GDP per capita and life expectancy, both in the short and long run. Their variance-decomposition results further indicate that demographic factors account for a substantial share of healthcare spending fluctuations, while feedback effects from expenditure to demographic and economic variables remain limited.
Micro-level evidence provides more nuanced results. Using individual-level data on primary care expenditure, \cite{atella2014effect} find that age is a stronger predictor of healthcare costs than proximity to death, challenging the “red herring” hypothesis for the Italian case and contrasting with earlier findings by \cite{raitano2007determinants}. Other disaggregated analyses examine the impact of macroeconomic shocks, with \cite{sarti2017poverty} showing that the 2008 financial crisis reduced healthcare spending among poorer households compared to non-poor ones.
A complementary strand of literature reverses the causal perspective by analyzing the effects of healthcare expenditure on health outcomes. \cite{golinelli2017health} find that higher public spending on directly provided services significantly reduces short-term mortality, while expenditure financed by the National Health Service (NHS) but delivered by private providers shows no significant effect. Related studies examining post-crisis fiscal consolidation policies highlight heterogeneous effects: while recovery plans effectively reduced regional deficits without affecting utilization or health outcomes \citep{bordignon2020hardening}, other contributions document increases in avoidable mortality, reduced hospital capacity, and worsening mental health outcomes following austerity measures \citep{arca2020death,cirulli2023austerity}.
Recently, \citet{caravaggio2026determinants} used ML algorithms to identify the determinants of healthcare expenditure in Italy, confirming several of the major drivers already documented in the literature.

\subsection{Applications of Symbolic Regression}

SR has recently been applied in many fields of science and engineering, such as physics, material science, finance, and medicine, due to its inherent ability to retrieve interpretable models \citep{angelis2023artificial, dong2025recent}. The discovery of explicit expressions can be useful in several domains. For instance, \cite{rezaei2022seismic} applied SR to predict the seismic response of bridges, while \cite{yang2015modeling} employed SR to model oil production and predict future trends. \cite{lv2026hybrid} proposed a hybrid framework to forecast agricultural water-demand in which SR is used to extract interpretable predictive equations. \cite{pan2019influential} detected the most influential factors for carbon emission intensity in different OECD countries via SR. \cite{ferrari2022multi} proposed a multi-objective SR framework to support clinical decisions in the healthcare domain. \cite{fong2024symbolic} used SR to estimate a key indicator for kidney health, the Glomerular Filtration Rate, to support disease diagnosis. 
In the literature, many authors highlight the possibility of exploiting SR to support policy-makers, by extracting influential drivers to guide strategic decisions \citep{lv2026hybrid,mcdermott2025symbolic}. 
\cite{truscott2014explaining} claims that SR can offer valuable support for socio-economic sciences applications, e.g., for the prediction of employment rates. \cite{balla2025ai} supports the usefulness of SR to extract interpretable symbolic models capable of uncovering hidden non-linear and dynamical relationships within real-world social science data, especially economics and sociology. 
\cite{miyazaki2023application} demonstrates the benefit of SR in the psychology domain by evaluating it to discover a discount function for an inter-temporal choice experiment. \cite{mcdermott2025symbolic} created surrogate models through SR for the global energy-economy-climate system, thus obtaining interpretable expressions that estimate future decarbonization paths. \cite{yang2015modeling} modeled the relationship between the economic growth and carbon dioxide emission for $67$ countries, while \cite{liu2023exploratory} explored the use of SR to detect influential factors, such as GDP and population, of sulfur dioxide emission in Chinese provinces.  Finally, \cite{claveria2020economic} employed SR to retrieve country-specific confidence indicators from qualitative survey expectations on economic growth for both consumers and firms, retrieving interpretable expressions for GDP growth forecasting. 
Despite the increasing number of applications in the socio-economic field, to the best of our knowledge, no one has focused yet on the analysis of influential factors of healthcare expenditures via SR.

\subsection{The Risk of Data Leakage}

ML is typically applied following a standardized pipeline, from data processing to error analysis. However, practitioners often employ ML algorithms without paying much attention to the choices made and why the obtained model behaves in a certain way. In this regard, \textit{data leakage} has been identified as one of the main challenges encountered in ML applications. 
\cite{apicella2025don} defines data leakage as \textit{``a problem where forbidden information is unintentionally introduced into the training data, thus affecting model performance evaluation''}. \cite{kapoor2023leakage} investigated reproducibility issues caused by data leakage in many ML scientific applications. Their analysis shows how the inflated performance of the model related to data leakage can lead to erroneous scientific evidence claims, thus causing  unreproducible results. In their taxonomy, performance overestimation is due to several reasons, one of which is an unclear separation between training sets and test sets.
\cite{cerqua2025mis} discussed the challenges related to the use of ML algorithms on panel data. Their claim is that, when ignoring the cross-sectional and longitudinal structure of data, the phenomenon of \textit{data leakage} becomes hard to detect. In fact, while standard ML typically splits the original dataset in training and test sets randomly, this approach can be dangerous for certain types of data. In the case of panel data, random splits cause temporal and cross-sectional leakage. This phenomenon has to be taken into account as neglecting it can cause inflated performance that do not reflect the real predictive power of the model. 

The aforementioned socio-economic sciences applications of SR do not consider the temporal perspective of datasets, at most focusing on county- or region-specific experiments \citep{pan2019influential, balla2025ai, liu2023exploratory}. Although some studies perform out-of-sample forecasting experiments on latest data, such as \cite{claveria2020economic}, the problem of temporal data leakage is not explicitly treated.
Our goal in this contribution is to directly address this issue by adopting a methodology that carefully splits training and test sets to avoid inflated results, and creates temporal windows to train models over different time horizons. 

\section{Methodology}
\label{sec:methodology}

\subsection{Data}

The analysis is based primarily on administrative data supplied by the Italian National Institute of Statistics \citep{ISTAT-HfA-2023} through Health for All (HFA), a dedicated tool for health-related data (last update accessed: June 2025). This source consists of a comprehensive database of indicators on the Italian healthcare system and population health, structured to be queried via the HFA software developed by the World Health Organization (WHO), and adapted to national requirements. At present, the database includes approximately 4,000 indicators disaggregated at the regional level, which constitutes the level of stratification adopted in this study. The sole variable which does not come from HFA is that of CO2 emissions, retrieved from \cite{ISTAT-statistics-2023}.

The initial dataset employed here consists of a total of $32$ features, listed in Table \ref{tab:listOfFeatures}, selected by domain experts. These features are measured for all Italian regions, from 1997 to 2019. 
Since the purpose is to create a forecasting model for healthcare expenditure, and past observations can strongly influence what happens in the future, all the features were lagged by one and two years with respect to the target variable. That is, for each feature $x$ we generated lagged $x_{t-1}$ and $x_{t-2}$ features with respect to the target measured at year $t$. The SR algorithm will then select variables among these lagged features.

In order to analyze both the public and private aspects of healthcare spending, three different target variables were considered: private (out-of-pocket) healthcare expenditure ($he\_pri$); public healthcare expenditure ($he\_pub$); and total healthcare expenditure ($he\_tot$) (obtained by adding public spending to private spending). All these variables are expressed in current euros per capita. Three different datasets were then obtained, one for each target.\\
Table \ref{tab:descriptive_lag1_and_targets} shows some statistics for each target and input variable, while Figure \ref{fig:heatmap} represents the heatmap for Pearson correlation among features in the datasets.

\subsection{Data Leakage prevention}

The typical ML pipeline involves random splitting to obtain training and test sets from the initial dataset. However, as outlined before, in panel data settings this procedure causes data leakage. In our work, following the principles laid down by \cite{cerqua2025mis}, we focus on \textit{temporal leakage}, defined as the phenomenon that \textit{``occurs when information from the future leaks into the past during the training and validation process''}. As such, if this issue is ignored, temporal information gets mixed between the training and test sets. This causes the model to see nothing new in the testing phase, but rather to review patterns already learned in the training phase. \\
The initial dataset employed in this work gathers socio-economic features for Italian regions across several years. Since our goal is to propose an ML-based strategy that is robust, we introduce a training and test set splitting strategy into our pipeline that avoids temporal leakage, based on the following two steps: 
\begin{itemize}
    \item \textit{Time-wise} splitting: the model is trained on earlier years and evaluated only on a separate set of future years. This is the solution proposed for sequential forecasting policy problems, where historical data are intended to be used for future predictions \citep{cerqua2025mis}. In our case, the test set was built considering the 25\% of the available years in the dataset. So the training set contains years from $1997$ to $2013$, while the test set the remaining years, i.e., $2014$-$2019$;
    \item \textit{Time windows} splitting: the socio-economic nature of the dataset means that there is heterogeneity of information across different time periods. In fact, healthcare expenditure can depend on various factors that may differ over the years due to political, historical, economic, and other significant changes. These considerations have led us to propose a further subdivision of the dataset across different time horizons. Four splits were obtained from the initial dataset, again using the time-wise strategy, placing more recent years in the training set and future years in the test set. This can help us capture different dynamics and see if obtained models return different results depending on the years considered. Table \ref{tab:timewindows} shows the details of the proposed time windows. 
\end{itemize}

The subdivision in different time windows sought to take into account changes due to the 2007-2008 financial crisis, which may have influenced healthcare expenditure management policies. Three time windows were therefore generated, corresponding respectively to the periods before, during, and after the crisis, together with a fourth window covering more recent years.

\subsection{Symbolic Regression}

This paper aims to employ a SR algorithm to detect drivers of healthcare expenditure in Italian regions, focusing on avoiding data leakage and discovering interpretable expressions. SR has the unique ability to predict the explicit function that links the dependent variable and the independent variables, making it a powerful tool for interpretability. SR is a ML technique that aims to retrieve the mathematical expression that best fits the input data.
It differs from traditional regression techniques for its ability to simultaneously discover the structure and coefficients of the formula representing relationships within data, without the need for prior knowledge. 
Given a set of training data $D = \{X,y\}$, where $X = \{x_1, \dots, x_m\}$ with $X \in \mathbb{R}^{n\times m}$ is the set of features (independent variables) and $y \in \mathbb{R}^n$ is the regression target (dependent variable), the goal of SR is to find a function $\hat{f}(x_1, \dots, x_m) = \hat{y}$ which best approximates $y$, i.e., $y - \hat{y} \leq \epsilon$. Thus, SR allows to find a mathematical model that best approximates the relationship between input features and the target value, simultaneously estimating the functional structure and the variable coefficients. As such, the main advantage of SR is its intrinsic interpretability \citep{dong2025recent}. In fact, SR allows us to find symbolic expressions that make explicit the links between different features. Although lately deep learning has been used to develop SR methods, Genetic Programming-based SR algorithms still remain the predominant approach \citep{la2021contemporary}. Genetic Programming (GP), proposed by \cite{koza1994genetic}, is inspired by the Darwinian natural selection principle. A population of individuals (originally computer programs) is randomly generated and then evolved with mutation and crossover operators, iteratively adapting to the problem at hand. In the context of SR, individuals are represented by mathematical expressions. Iteration after iteration, the individuals, i.e., the expressions, that best represent the initial dataset, survive. The expressions in the initial population are built combining mathematical functions and operators, symbols and constants, representing the potential solution, i.e., the model which best fits the input data. At each iteration, new offspring of individuals are created by means of mutation and crossover. The mutation operator randomly changes the parent structure. The crossover operator acts as reproduction between two individuals, by exchanging parts of the parents' expressions. Figure \ref{fig:mutcross} shows an example of mutation and crossover operators. The selection of the expressions is performed with a \textit{fitness metric}, which measures how well the expressions of the current offspring represent the dataset.

In this paper, we employed Operon \citep{burlacu2020operon}, a GP-based C++ framework for SR, which has been ranked as one of the best methods in SRBench++ benchmark \citep{imai2025call,la2021contemporary}. 
In this work, we chose to use SR because of its ability to provide explicit and interpretable expressions. This makes it possible to study the analytical link between the factors that influence public and private healthcare spending. The resulting expressions can be useful tools for intuitively visualizing linear and nonlinear relationships, signs, coefficient values, and the overall structure of the function. Furthermore, the importance of variables can be studied. First, SR acts as an automatic feature selection tool, including in the final expressions only those variables considered as important. Secondly, the stability of the selection of these variables can be analyzed over multiple runs of the same experiment. In fact, for each dataset and splitting strategy, the SR algorithm is run $100$ times, thus obtaining $100$ expressions. By counting the occurrence of each feature in these expressions, it is possible to assess which are the most frequent variables, which are those that appear most consistently in different expressions, presumably suggesting that they are the most important \citep{kronberger2024symbolic}. However, this is an algorithm-dependent selection measure, rather than a direct measure of causal importance, statistical significance, or an intervention effect. Later, we also explored the frequency of functional substructures, which might be indicative for exploring multi-variable interactions. 

For each dataset (corresponding to $he\_pri$, $he\_pub$, or $he\_tot$ targets), the SR algorithm was applied separately for each splitting strategy, in order to analyze their effects on the final results. It is worth saying that the algorithm is trained on standardized features and targets, thus the coefficients of the resulting expressions refer to standardized inputs and the output value does not correspond to the original unit. 
Two distinct metrics were taken into account for models evaluation: the coefficient of determination $R^2$ and the Mean Percentage Error (\textit{MAPE}). $R^2$ is one of the most commonly used error metrics for SR models assessment \citep{la2021contemporary,tian2025interactive, dong2025recent}, measuring the goodness of fit of resulting models. However, using $R^2$ values alone is not entirely indicative of the accuracy of results. As such, we decided to consider \textit{MAPE}, recently used to evaluate SR models \citep{panczyk2025opening,megahed2025symbolic}, to quantify prediction errors.


\section{Results}
\label{sec:results}


\subsection{Experimental settings}
The SR algorithm was executed for a total of $100$ runs for each dataset and splitting strategy (time-wise and temporal windows $W_1$, $W_2$, $W_3$, $W_4$). This was done in order to obtain as much variability as possible in the results, and thus assess the robustness of the proposed method. 
The general settings of the algorithm parameters were based on the results of the tuned SR methods shown by SRBench++ \citep{la2021contemporary}. After several trials, we decided to maintain the following setting: (i) $\text{function set}=\{+,-,*,/,sqrt\}$, as adding more complex operators (such as logarithmic or exponential functions) demonstrated lower performance and displayed less interpretable expressions; (ii) the objective was maintained to $R^2$, as using other metrics or adding complexity constraint did not improve final results; (iii) the maximum depth and length were both maintained to $10$, as their increase generated over-complicated expressions without improving fitness, thus causing bloating issues \citep{kronberger2024symbolic}.

\subsection{Healthcare expenditures models}

We now show the results obtained by running SR on the three different datasets $he\_pri$, $he\_pub$, and $he\_tot$.
Even if the random splitting strategy for panel data is not advisable, we still executed it to demonstrate the presence of data leakage. 
Figure \ref{fig:randomtimewise} shows inflated performance in $R^2$ values when using random splitting strategy. While the $R^2$ for private health expenditure ($he\_pri$) remains good ($76\%$), results on public health expenditure ($he\_pub$) drop significantly ($20\%$). Instead, \textit{MAPE} decreases, showing better predictions for time-wise settings. These results confirm the claim that splitting panel data randomly is not a reliable strategy.  For this reason, we focus on results obtained with time-wise and time windows splittings. The values of the average $R^2$ and \textit{MAPE} scores are summarized in Figure \ref{fig:timewisewindows}. 
$R^2$ score values for $he\_pub$ remain low also for temporal windows splittings, while \textit{MAPE} remains under $10\%$ which according to \cite{lewis1982industrial} is the threshold for highly accurate forecasting.
This means that the models return predicted values close to the real ones on average, but fail to capture the dynamics of the phenomenon. 
Time window $W_4$ has the best results in terms of both $R^2$ (for private and total healthcare spending), and \textit{MAPE}. Instead, $W_3$ shows the lowest $R^2$ average scores and the highest \textit{MAPE} average scores, especially for private spending, where \textit{MAPE} is greater than $10\%$. This means that SR was able to identify relevant drivers of the phenomenon and the relationships between them for recent years ($W_4$ is trained and tested on the latest time horizon), while it failed to obtain good models for $W_3$. The latter considers a time period characterized by the $2008$ financial crisis, which probably had an impact on the healthcare spending as well as drivers guiding it. 
In light of the results obtained for public and total spending, we will now focus our analyses on private spending, a sector which shows more stable and reliable results.

\subsubsection{Private healthcare expenditure models}

As introduced earlier, the SR intrinsically performs feature selection by including in the solutions a subset of the input variables. Since the SR algorithm was run $100$ times over the same dataset, this gives us the opportunity to analyze the frequencies of variables in the obtained expressions. As stated before, the occurrences of features in different expressions for the same data could suggest their importance. 
Therefore, we make the following assumption: if a variable is consistently chosen across different runs on the same configuration (algorithm parameters, dataset and splitting strategy), then it might be a relevant driver for describing the dynamics of the phenomenon, but does not automatically demonstrate that it has a greater effect on expenditure.
Table \ref{tab:freqhepri} shows the occurrences of the top-5 most frequent variables appearing in the models for $he\_pri$. These frequencies were obtained by counting the number of times a variable appears in distinct resulting expressions, thus the unique occurrences within each model.

The variable $he\_pub\_p$, representing the share of public health expenditure, is nearly always the most recurrent predictor across split types. Its first lag is selected $100$ times in $W_3$ and $W_4$ and remains dominant in $W_2$ and time-wise splittings; its second lag appears 95 times in $W_1$.
The $a\_hh\_size$ variable, i.e., the average household size, also seems to be relevant, especially for $W_2$ and $W_3$, while it appears with less frequency \(W_4\). Besides these two dominant features, technological-related variables ($CT\_scan$, i.e., CT scanners per 1,000 inhabitants, and $MRI\_scan$, i.e., MRI scanners per 1,000 inhabitants) have a relevant role, even though with less strength when compared to $he\_pub\_p$ and $a\_hh\_size$. Features related to the labour market ($act\_rate$ and $empl\_rate$) appear less but still consistently. 

The analysis shifts then to the explicit expressions produced by the SR.
Since displaying all 100 expressions for each split is impractical, we report a small set of high-performing expressions to provide qualitative insight into the types of analytical relationships recovered by SR. These equations are not intended as final deployable forecasting models, but as examples of the broader family of solutions generated by the method.
Table \ref{tab:best-models} shows $3$ high-performing models obtained for each splitting strategy ordered by $R^2$ and $MAPE$ values. 
Results for $W_3$ were excluded for the poor performance of obtained models (see Figure \ref{fig:timewisewindows}) 

Time-wise and $W_1$ splittings display linear models, while $W_2$ and $W_4$ present some non-linearities, suggesting more complex dynamics in specific temporal settings. 
All these models contain the variable $he\_pub\_p$, confirming its strong role in the private spending phenomenon, with a negative sign and similar coefficients.
The GDP per capita is consistently present in all models, except for one expression for $W_4$.
$MRI\_scan$, one of the technological-related drivers, is only absent in $W_4$, and appears as a non-linear contribution in $W_2$. 
It is worth noting that some variables appearing in these illustrative expressions do not belong to the top-5 most frequent features discussed above (see Table \ref{tab:freqhepri}). For instance, $h\_beds_{t-2}$, representing the number of hospital beds per 1{,}000 inhabitants, appears in all three reported time-wise expressions despite having a relatively low overall frequency of 7 out of 100 runs. Similarly, $p\_r\_75ov\_p_{t-2}$ occurs in two of the three reported expressions for $W_1$, although it is not included among the five most frequent variables for that split and appears in 27 out of 100 runs overall. This suggests that frequency alone does not fully capture the potential relevance of a variable: some predictors may not be among the most recurrent across runs, yet still contribute to high-performing symbolic forms. More generally, these results indicate that SR does not return a single canonical equation, but rather a family of candidate analytical relationships, within which both highly recurrent variables and more occasional predictors may play a meaningful role.

\paragraph{Out-of-time models evaluation}

We assessed the generalization of the resulting models through their out-of-time evaluation on new temporal data, i.e., 2020 and 2021, which were not used during either model training or testing. The aim here is to understand to what extent the relationships recovered by SR remain informative when transferred to another temporal context. Tables \ref{tab:eval-pri-2020} and \ref{tab:eval-pri-2021} show comparisons of $R^2$ and \textit{MAPE} values between results on the testing set and this evaluation set. 
The most recent temporal window $W_4$ is the one for which the resulting models show the best temporal transfer to 2020. This is reasonable as these expressions have been trained over the closer years, thus better capturing recent expenditure dynamics. In fact, it is the only splitting in which the out-of-time evaluation for year $2020$ returns good results in terms of both $R^2$ and \textit{MAPE}. 
By contrast, results strongly degrade for $2021$, including $W_4$. Although this period coincides with major changes associated with the COVID-19 pandemic, the present analysis cannot determine whether the decline in performance was caused by the pandemic.

We also evaluated on 2020 the three illustrative high-performing expressions previously reported for the time-wise split and for $W_4$.
The other time windows were excluded from this analysis due to their weaker behaviors over out-of-time data. The time-wise models returned $R^2$ values of $0.73$ for the first expression and $0.77$ for the other two, with \textit{MAPE} values around $7\%$.
The three $W_4$ expressions obtained $R^2$ values of $0.83$, $0.83$ and $0.73$, with corresponding \textit{MAPE} values of $5.80\%$, $7.48\%$, and $8.78\%$.
These results further confirm that expressions performing well on the test set are not necessarily those that generalize best to later unseen years. This is consistent with the nature of healthcare expenditure as a complex and evolving socio-economic phenomenon, whose relevant drivers and functional relationships may vary across time horizons.
For this reason, we further analyzed the expressions obtained over $W_4$ temporal window for the $2020$ data, since this split represents the most recent time period and its results over out-of-time data are the most convincing. Table \ref{tab:best-models-eval} summarizes the top-10 best-performing expressions for $W_4$ on the $2020$ out-of-time evaluation ranked by $R^2$. 

These results suggest frequent and consistent substructures in the expressions with highest performance. To explore this hypothesis, we conducted a further analysis on the expressions obtained via SR to study the frequency of each functional structure and substructure. Formally, we define the \textit{structure} of an expression as its symbolic skeleton, where numerical coefficients are mapped to a generic placeholder, e.g., the structure of $2*a + 3*b + 4*c $ is $C*a+C*b+C*c$. A \textit{substructure} is defined as any valid sub-tree or combination of nodes derived from the expression tree. Following the previous example, the substructures of $C*a + C* b + C*c$ include individual terms like $C*a$, as well as partial combinations such as $C*a + C*b$ or $C*a + C*c$. This approach allows us to isolate the recurrent building blocks in the retrieveed expressions. We show here the results of this analysis for the $100$ expressions of $W_4$ temporal window and substructures with at least two variables (as considering one-variable substructure is equivalent to counting variable frequency).  Table \ref{tab:freq-substructures} shows the top-$15$ substructures for $W_4$ ordered by their frequency, i.e., the number of times that substructure appears in distinct expressions. We can see that the interactions of the share of public health expenditure ($he\_pub\_p$) with the foreign population and the average household size are the most frequent. It is interesting to observe the substructure of $ C*a\_hh\_size_{t-2} + C*he\_pub\_p_{t-1} + C*empl\_rate_{t-2} + C*str\_r\_p_{t-2}$, composing the 17\% of the resulting expressions, is present in $7$ out of the best $10$ expressions obtained for $2020$, while the first two expressions include the same substructure but with $a\_hh\_size_{t-1}$ instead.
This analysis enriches the previous results in variable frequencies. Although the occurrence of a single variable can be indicative of its importance in describing the phenomenon at hand, the frequency of the interaction between different features highlights which functional substructure appears the most. However, these are recurring joint predictive patterns which should not be directly considered as interaction terms. The recurrence alone does not identify an underlying mechanism or establish that a substructure improves performance relative to an otherwise comparable expression without it. Nevertheless, we still believe that these analyses highlight structural combinations of terms that are worth to be investigated by domain experts.

\section{Discussion}
\label{sec:discussion}

In this work we investigated the use of SR to provide interpretable expressions to better understand the role of potential determinants for healthcare expenditure in Italian regions. The splitting strategy adopted for the input dataset specifically addresses temporal data leakage, in order to avoid inflated performance. A comparison between random and time-wise splitting has also been made. Results show that the latter strategy leads to smaller $R^2$ values, demonstrating that training the algorithm on randomly sampled data causes data leakage. This is due to the fact that random sampling mixes future and past information across training and test data, causing the algorithm to see nothing new in the test phase, as it has already learned future patterns in the training stage. Instead, time-wise splitting carefully separates time horizons, making the algorithm actually test on unseen temporal data. The effects of this phenomenon are particularly relevant for the public health expenditure, with $R^2$ decreasing from $0.85$ to $0.20$.
However, even if time-wise splitting is useful to reduce data leakage, in the case of complex socio-economic phenomena such as healthcare spending, training over a large time interval that comprehends several political and historical changes may not be effective to obtain reliable and robust analytical models. Therefore, we proposed a splitting strategy including different temporal windows. Results showed that SR returned expressions with different structures, features, and performances for each temporal window, thus confirming our claims. In fact, in Figure \ref{fig:timewisewindows} we can observe visible differences in error scores between different splitting strategies. The temporal window $W_3$ gives the worst performance overall, especially for the private expenditure in which the average \textit{MAPE} reaches the $15.57\%$. The training set corresponding to $W_3$ includes years $2001$ - $2009$, while the model has been tested on years $2010$ and $2011$. The test set corresponds to a particularly relevant historical time, as it corresponds to the financial post-crisis period. SR struggled to find accurate models for this time window, probably due to the impact of the economic and political instability on the drivers of healthcare spending \citep{arca2020death,bordignon2020hardening}.

For each splitting strategy, we also analyzed the frequency of variables appearing in the resulting expressions for the private healthcare expenditure. Using variable frequency as an algorithm-dependent measure of recurrence rather than of causal influence, the models obtained across temporal windows highlight different sets of recurrent predictors for private healthcare expenditure (see Table \ref{tab:freqhepri}). The use of multiple temporal splits allows to detect predictors that are consistent across different time horizons (\textit{e.g.}, the share of public health expenditure), as well as those that are relevant for specific time periods (\textit{e.g.}, the employment rate, which is relevant only for the $W_4$ temporal window). Across splitting strategies, the models evolve from relatively simple, macro-driven specifications to more complex representations centered on health system dynamics. While income and public expenditure remain the structural backbone, later models highlight that out-of-pocket spending's models contain interactions between policy, workforce, and system characteristics. 

A clear divergence in predictive performance emerges between private and public healthcare expenditure. While random splitting initially masked this difference - yielding models with average $R^2$ values around 0.85 - the adoption of time- and window-based splits reveals the intrinsic difficulty of forecasting public health spending. Using the same set of explanatory variables, the SR algorithm consistently struggles to identify stable and accurate relationships for public expenditure, with model performance not exceeding an $R^2$ of 0.42. In contrast, models for private out-of-pocket expenditure achieve substantially higher predictive accuracy.
This asymmetry can be largely explained by the institutional design of the Italian National Health Service (NHS). Public healthcare expenditure is not primarily determined by market dynamics or short-term variations in population needs, but rather by a structured and rule-based allocation process. In particular, the bulk of funding is distributed through a capitation formula applied to the undifferentiated financing component, which represents the largest share of total resources. Although this formula incorporates some demographic and need-based adjustments, it evolves slowly over time and is only partially captured by the variables included in the model. Additional funding components - such as earmarked resources and the premial quota - are allocated through discretionary and negotiated processes within the State-Regions Conference \citep{caruso2019salute}.
Moreover, following the period of rapid expenditure growth and fiscal imbalances in the early 2000s, several Italian regions were placed under strict financial recovery plans. These regimes impose binding budget constraints, enforce expenditure ceilings, and promote convergence toward predefined fiscal targets. As a result, public healthcare spending exhibits relatively smooth and constrained dynamics, with adjustments driven more by policy decisions and administrative controls than by observable socio-economic or epidemiological factors. From a modeling perspective, this reduces the effective variability in the data and weakens the statistical relationship between expenditure and standard predictors, making it difficult for SR to uncover meaningful functional forms.
By contrast, private health expenditure is inherently more responsive to individual behavior and market conditions. It reflects households’ decisions to seek care outside the public system, often influenced by income, demographic structure, perceived quality of services, waiting times, and availability of medical technology. These determinants are more directly reflected by the variables included in the model (such as GDP per capita, healthcare capacity, and utilization indicators), resulting in stronger and more stable empirical relationships. Additionally, private expenditure is not subject to the same institutional rigidities as public expenditure and therefore exhibits greater variability, which provides a richer signal for data-driven methods like symbolic regression.
Overall, the contrast in predictive performance highlights a fundamental distinction: public healthcare expenditure in Italy is largely institutionally determined and only partially observable through standard covariates, whereas private expenditure is more demand-driven and closely aligned with measurable socio-economic and healthcare system factors. This difference naturally translates into a substantially higher predictive accuracy for private healthcare expenditure models within the SR method, as it was able to find good models for private health expenditure, especially for $W_4$ time window in which the average $R^2$ is equal to $0.92$ (an excerpt of the resulting analytical models being shown in Table \ref{tab:best-models}). 

While the study on the frequencies of variables highlighted different relevant drivers for each time window, the analysis of the resulting expression indicates different functions and relations between variables for each time window. This demonstrates that the considered time period has a strong impact on the form of the models obtained as well as the selected features. The variable $gdp\_cap$, as expected, is included in all model specifications and consistently displays a positive coefficient. This is reasonable, as it proxies household well-being and their ability to rely on co-financing (\textit{i.e.}, ticket payments) and out-of-pocket healthcare expenditure. 
The presence of high-cost medical equipment within each region, such as CT and MRI scanners ($CT\_scans$ and $MRI\_scans$, respectively), appears in all model specifications except for the last one, which corresponds to the years characterized by financial constraints within the Italian National Health Service (NHS) \citep{arca2020death}. In the literature, the availability of such equipment is generally considered a proxy for technological advancement, which, in the original formulation of \citep{newhouse1977medical}, explains the residual component of healthcare expenditure. This mechanism undoubtedly applies to the private sector (both accredited and fully private providers), which must attract private demand by offering not only faster access but also higher quality services, particularly when the public sector falls short.
The variable referring to the share of public healthcare expenditure ($he\_pub\_p$) over total expenditure is consistently negatively associated with our target variable. However, the result does not identify substitution or crowding out between public and private healthcare. Similarly, the number of beds in public and accredited facilities follows the same pattern, although this variable appears less consistently across the different model specifications. It is noteworthy that, in the $W_{4}$ specification, the share of public expenditure exhibits a less linear relationship with the dependent variable. This pattern likely reflects the period of fiscal imbalances and the placement of several Italian regions under government-appointed health commissioner administration.
In some specifications, specifically the Time-Wise and $W_{2}$ models, the short-run equations show a negative association between average family size and the dependent variable. 
This is compatible with hypotheses involving household composition and informal care discussed in \citep{marmot2005social,dormont2006health, cisotto2023informal}, but the available analysis cannot identify a household scale effect or a caregiving mechanism. However, this effect disappears in subsequent model specifications referring to more recent time horizons. This change likely reflects the increasing propensity of Italian households - already experiencing sustained demographic decline - to delegate primary caregiving responsibilities to external (privately paid) care workers, including for home care services \citep{lippi2016delegating}.
Finally, the share of the population aged 75 and over ($p\_r\_75ov\_p$) exhibits a positive association, confirming previous findings reported by the literature on the determinants of healthcare expenditure \citep{gerdtham1992econometric,giannoni2002regional}.

In order to evaluate the generalizability of the models to unseen years, we applied models obtained for private spending in $2020$ and $2021$ years (Tables \ref{tab:eval-pri-2020}, \ref{tab:eval-pri-2021}). The results show that the models obtained with the most recent temporal window $W_4$ performed the best in $2020$. Despite the decrease in accuracy, the resulting $R^2$ and \textit{MAPE} values still make the models usable. It is a different case for $2021$, where performance decreases significantly, with error values indicating useless models. The pandemic literature \citep{trentini2023repeated,coccia2025dynamics} provides relevant context, but this study cannot attribute the deterioration specifically to COVID-19. 
In addition, we assessed the best-performing models obtained for $2020$ data based on $W_4$ expressions. The results in Table \ref{tab:best-models-eval}, displaying the top-$10$ analytical models, indicate consistent structures and interactions between variables. In fact, the obtained models nearly always contain interactions between the average household size ($a\_hh\_size$), the share of public health expenditure ($he\_pub\_p$), the employment rate ($empl\_rate$), and the foreign population ($str\_r\_p$). This led us to conduct an additional analysis in which we studied the frequency of substructures obtained among the $100$ runs of SR. Similarly to the frequency of variables, we can assume that if a substructure consistently appears in different runs for the same time window, it may represent a relevant relationship describing the phenomenon. For example, from Table \ref{tab:freq-substructures} we can state that the additive terms involving $he\_pub\_p$ and $str\_r\_p$, and  $he\_pub\_p$ and $a\_hh\_size$ are the most common. 

To summarize, our goal was not to provide a unique best expression that describes the investigated phenomenon. In this regard, GP-based SR methods should be considered as \textit{hypotheses generating machines} \citep{kronberger2024symbolic}, given their ability to return multiple models with similar accuracy for the same problem. Especially for complex phenomena like the one analyzed here, SR allows to return different interactions between variables that describe different aspects of the same dimension. This should not be seen as a flaw, because it contributes to analyze the healthcare expenditure from different perspectives, via the study of variables' and structures' frequencies, thus providing interesting insights to assess the most relevant and influential interactions explaining the underlying dynamics of this phenomenon.\\

\section{Conclusion}
\label{sec:conclusion}

Identifying possible drivers in the dynamics of healthcare expenditure is essential to guide policymakers in strategic decisions, including the allocation of resources. This study investigates the use of SR to obtain interpretable analytical expressions in temporal constraint settings. In order to overcome temporal data leakage issues in panel data, we performed time-wise splitting for training and test data. Additionally, given the complexity of the phenomenon studied, we decided to consider different temporal windows, with results showing different expressions and recurrent variables depending on the time horizon.
SR has demonstrated to be effective in retrieving formulas with good accuracy for private healthcare expenditure, especially for the most recent temporal window.
By contrast, public healthcare expenditure remained harder to model, with lower predictive performance. This suggests that public spending is likely influenced by institutional and contextual factors that are only partially represented in the dataset, thus the available variables are not sufficient to model this phenomenon.
Although we retrieved high-performing formulas for private spending, our goal was not to pick out the best of all, but rather to exploit symbolic structures for the explicit analysis of relationships between input features, following the principle of SR as an hypotheses generating machine. The multiple runs performed over the same dataset and splitting strategy have been useful to study the robustness of the approach and the recurrence of features and substructures. The findings indicated that SR can provide useful information not only on the relevance of variables but also on how they combine within repeated analytical patterns.
However, although SR has proved beneficial for deriving interesting insights from the generated expressions, this work has some limitations. Firstly, the relevance of the candidate drivers is based on how often the SR models include them in the final expressions, and is therefore not supported by additional causal analyses. Indeed, frequency-based analysis is intended to be exploratory rather than an established methodology for causal identification. Furthermore, the fact that the models vary significantly from one time horizon to another and that their out-of-time performance deteriorates markedly for 2021 indicates their instability in the face of economic shocks, thereby limiting their applicability for long-term analysis. Another limitation concerns the SR algorithm used: GP-based methods are highly sensitive to their hyperparameters and do not always explore the search space efficiently. It is therefore worth exploring the use of alternative SR paradigms, such as those based on Large Language Models, and to assess whether the recovered variables and substructures remain stable across different algorithms.

Overall, the paper shows that SR is a promising tool for the interpretable analysis of socio-economic phenomena, providing expert end-users a way to investigate explicit relationships between drivers of interest. At the same time, the results make clear that model quality depends strongly on the temporal setting, on the target variable considered, and on the availability of informative predictors. Future works should consider enriching the set of features, exploring more in-depth the analysis of recurrent substructures, and further testing the robustness of the recovered relationships on post-pandemic years and other territorial or national contexts.

\section*{Statements and Declarations}

\subsection*{Funding}
No funds, grants, or other support were received for conducting this study or preparing this manuscript.

\subsection*{Competing interests}
The authors have no relevant financial or non-financial interests to disclose.

\subsection*{Ethics approval and consent to participate}
This study used aggregated regional administrative data and did not involve human participants, human material, or animals. Ethics approval and informed consent to participate were therefore not required.

\subsection*{Consent for publication}
Not applicable.

\subsection*{Data availability}
The processed data supporting the findings of this study are available from the corresponding author upon reasonable request. The underlying source data were obtained from the Italian National Institute of Statistics (Istat) Health for All database and Istat statistics, subject to the applicable terms of use.

\subsection*{Materials availability}
Materials supporting the findings of this study are available from the corresponding author upon reasonable request.

\subsection*{Code availability}
The code used to conduct data processing and the reported analyses is available from the corresponding author upon reasonable request.

\subsection*{Author contributions}
All the authors conceived the study. Cristina Rossetti and Nicola Caravaggio developed the theoretical framework and drafted the manuscript. Nicola Caravaggio performed data collection and curation. Cristina Rossetti performed data cleaning and carried out the experiments. Giovanni Lamura and Giuliano Resce supervised the project and critically revised the final version of the manuscript. All authors discussed the results, contributed to the final manuscript, and approved the submitted version.

\pagebreak

\bibliographystyle{apalike-etal-in-italics} 
\bibliography{bib}

\pagebreak

\section*{Appendix}
\label{section:appendix} 

\setcounter{section}{0}
\setcounter{subsection}{0}
\setcounter{subsubsection}{0}
\renewcommand{\thesubsection}{A.\arabic{subsection}}
\renewcommand{\thesubsubsection}{A.\arabic{subsection}.\arabic{subsubsection}}
\setcounter{table}{0}
\renewcommand{\thetable}{A\arabic{table}}
\setcounter{figure}{0}
\renewcommand{\thefigure}{A\arabic{figure}}

\begin{figure}[H]
    \centering
\includegraphics[width=0.8\linewidth]{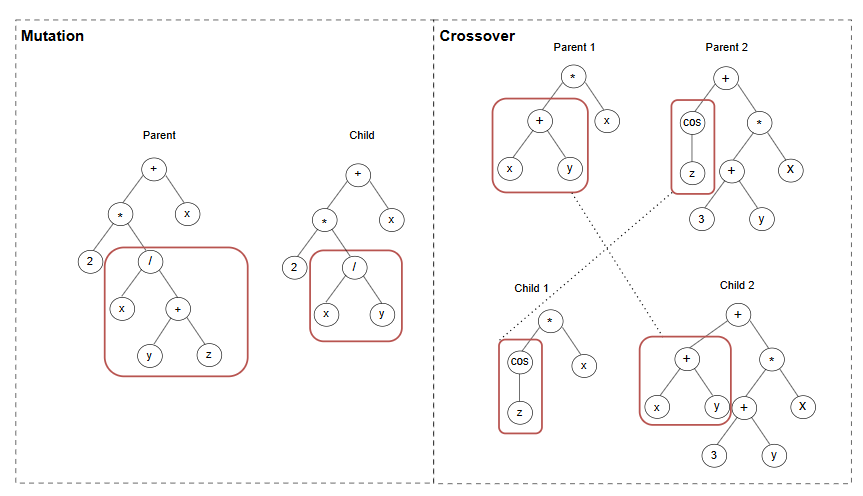}
    \caption{Mutation and crossover examples in Genetic Programming for Symbolic Regression. Here, the mutation operator transforms the parent expression ($x+2*\frac{x}{y+z}$) into its corresponding child ($x+2*\frac{x}{y}$) by replacing the substructure $\frac{x}{y+z}$ with $\frac{x}{y}$. The crossover operator takes two parents ($x*(x+y)$ and $cos(z) + x*(3+y)$) and recombines one of their substructure to create two different child expressions ($cos(z)*x$ and $(x+y) + x*(3+y)$). }
    \label{fig:mutcross}
\end{figure}

\begin{figure}[H] 
\centering \includegraphics[width=1.0\linewidth]{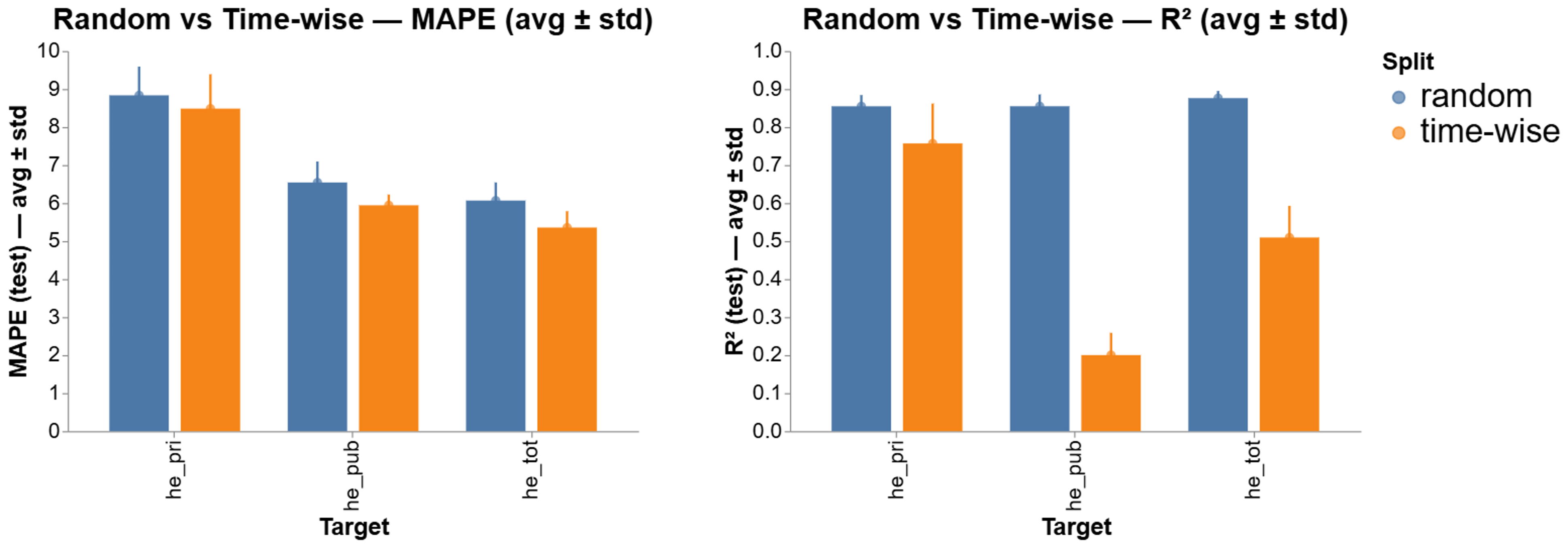} 
\caption{Average $R^2$ and \textit{MAPE} for random and time-wise splittings} \label{fig:randomtimewise} 
\end{figure}

\begin{figure}[H] 
\centering \includegraphics[width=1.0\linewidth]{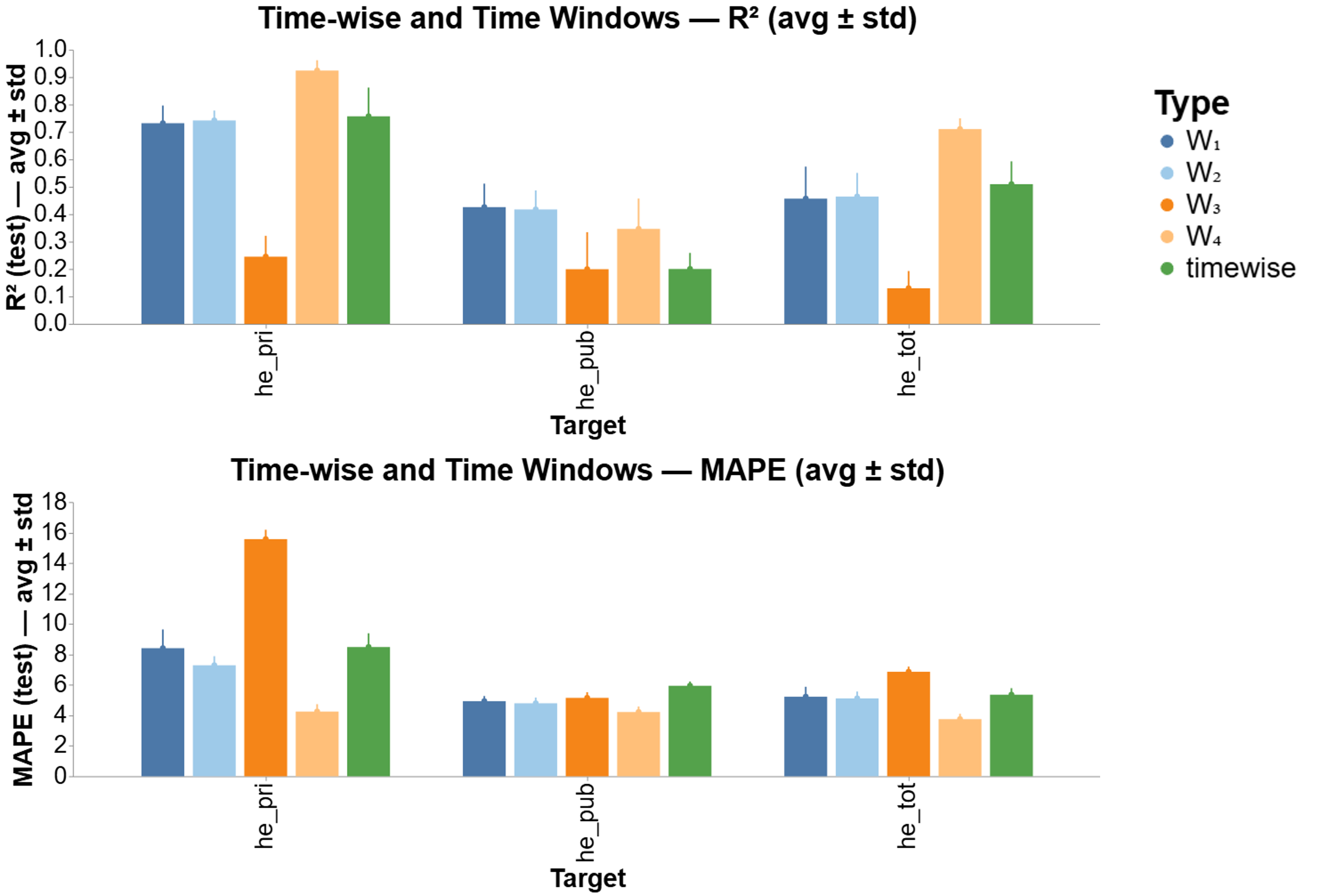} 
\caption{Average $R^2$ and \textit{MAPE} for time-wise and time windows splittings} \label{fig:timewisewindows} 
\end{figure}

\begin{table}[H]
    \centering
    \caption{List of dataset features}
    \label{tab:listOfFeatures}
    \begin{tabular}{ll}
        \hline
        \textbf{Feature} & \textbf{Meaning} \\
        \hline
        area & Region surface ($Km^2$) \\ 
        he\_pub\_p & Share of public health expenditure \\ 
        pop\_res & Resident population \\ 
        dens & Population density \\ 
        p\_r\_04\_p & Population aged 0-4 (\% of total) \\ 
        p\_r\_65ov\_p & Population aged 65+ (\% of total) \\ 
        p\_r\_75ov\_p & Population aged 75+ (\% of total) \\ 
        age\_ind & Ageing index \\ 
        p\_r\_m\_p & Male population (\% of total) \\ 
        p\_r\_f\_p &  Female population (\% of total)\\ 
        str\_r\_p & Foreign population (\% of total) \\ 
        a\_hh\_size & Average household size  \\ 
        fert\_rate & Fertility rate \\ 
        birth\_rate & Birth rate \\ 
        mort\_rate & Mortality rate \\ 
        act\_rate & Activity rate (15+) \\ 
        empl\_rate & Employment rate (15+) \\
        unempl\_rate & Unemployment rate (15+) \\ 
        obe\_p &  Obese individuals 15+ (\% of total)\\ 
        obe\_65ov\_p&  Obese individuals 65+ (\% of total)  \\ 
        ovweight\_p &  Overweight individuals 15+ (\% of total) \\ 
        ovweight\_65ov\_p & Overweight individuals 65+ (\% of total) \\ 
        smok\_p & Smokers 18+ (\% of total) \\ 
        smok\_65ov\_p &  Smokers 65+ (\% of total)\\ 
        big\_smok\_p & Heavy smokers 19+ (\% of total)  \\ 
        gdp\_cap & GDP per capita at current prices (€)  \\ 
        co2\_cap & $CO_2$ emissions per capita\\ 
        h\_beds & Hospital beds per 1,000 inhabitants \\ 
        physicians &  Physicians per 1,000 inhabitants\\ 
        nurses & Nurses per 1,000 inhabitants\\ 
        CT\_scan &  CT scanners per 1,000 inhabitants\\ 
        MRI\_scan & MRI scanners per 1,000 inhabitants \\ 
        \hline
    \end{tabular}
\end{table}

\begin{figure}[h]
    \centering
    \includegraphics[width=1.0\linewidth]{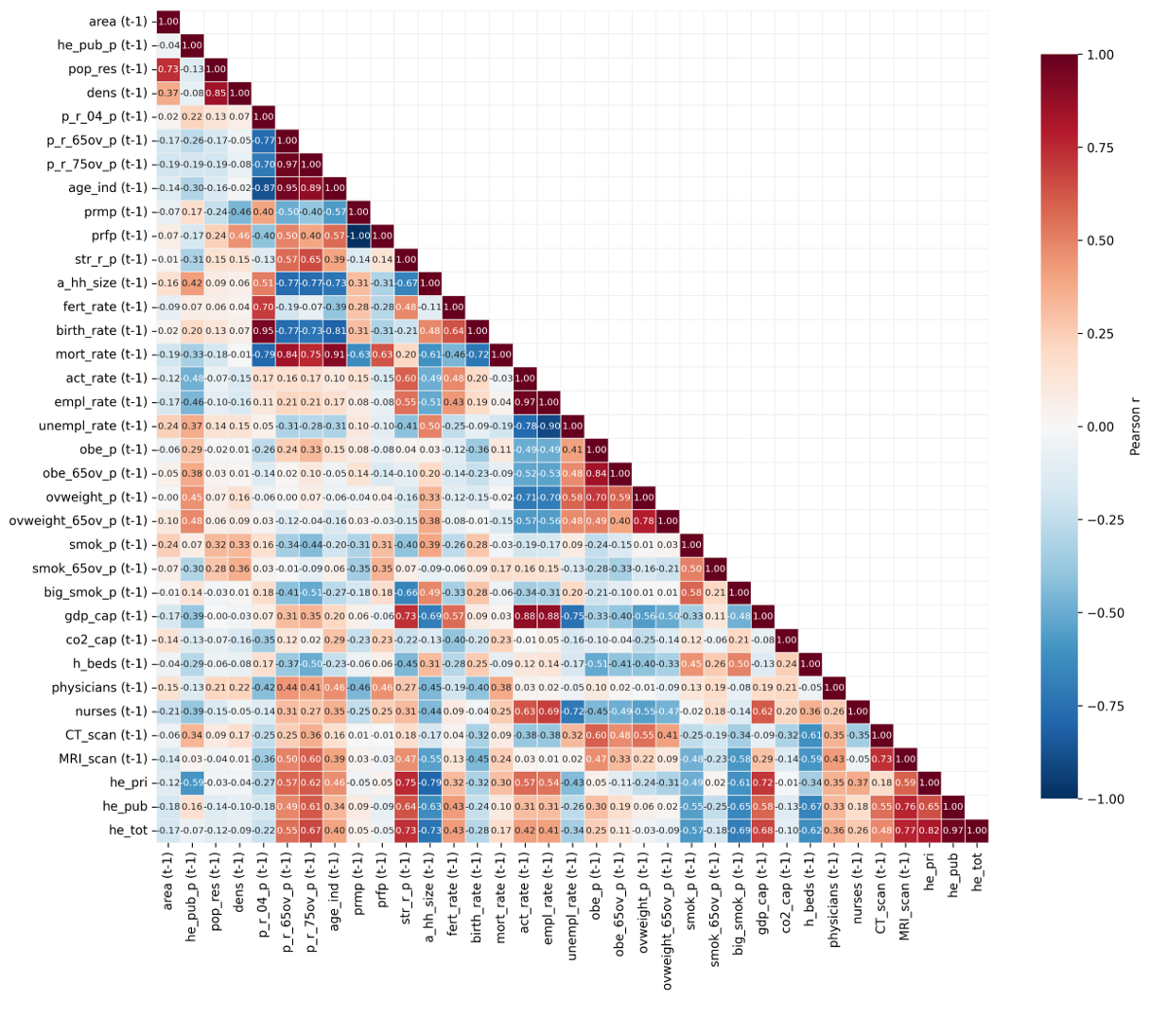}
    \caption{Heatmap showing Pearson correlation between features in the datasets.}
    \label{fig:heatmap}
\end{figure}

\begin{table}[H]
\centering
\scriptsize
\setlength{\tabcolsep}{2.5pt}
\caption{Descriptive statistics of predictors at $t-1$ and expenditure at $t$}
\label{tab:descriptive_lag1_and_targets}

\begin{tabular*}{\linewidth}{@{\extracolsep{\fill}}lrrrrrr@{}}

\textbf{Variable}  & \textbf{Mean} & \textbf{SD} & \textbf{Median} & \textbf{Min} & \textbf{Max} \\
\midrule
area ($t-1$)  & 14,348.58 & 7,315.25 & 13,637.64 & 3,260.85 & 25,832.55 \\
he\_pub\_p ($t-1$)  & 76.84 & 3.85 & 76.91 & 65.79 & 85.46 \\
pop\_res ($t-1$) & 2,718,406.48 & 2,358,489.36 & 1,625,306.50 & 117,329.00 & 10,053,264.00 \\
dens ($t-1$)  & 169.36 & 106.98 & 152.67 & 35.98 & 443.98 \\
p\_r\_04\_p ($t-1$) & 4.47 & 0.60 & 4.40 & 3.12 & 6.49 \\
p\_r\_65ov\_p ($t-1$) & 20.55 & 3.07 & 20.73 & 12.73 & 28.65 \\
p\_r\_75ov\_p ($t-1$)  & 9.91 & 2.12 & 9.93 & 4.62 & 15.86 \\
age\_ind ($t-1$)  & 152.65 & 38.78 & 150.80 & 64.38 & 259.84 \\
prmp ($t-1$)  & 48.57 & 0.44 & 48.60 & 47.15 & 49.48 \\
prfp ($t-1$) & 51.43 & 0.44 & 51.40 & 50.52 & 52.85 \\
str\_r\_p ($t-1$)  & 4.83 & 3.34 & 3.75 & 0.31 & 12.06 \\
a\_hh\_size ($t-1$)  & 2.51 & 0.23 & 2.48 & 1.91 & 3.12 \\
fert\_rate ($t-1$)  & 1,302.37 & 155.74 & 1,291.50 & 959.00 & 1,756.73 \\
birth\_rate ($t-1$)  & 8.77 & 1.31 & 8.62 & 5.40 & 12.32 \\
mort\_rate ($t-1$)  & 101.15 & 14.21 & 99.41 & 74.51 & 144.73 \\
act\_rate ($t-1$)  & 49.18 & 5.28 & 50.08 & 37.87 & 60.63 \\
empl\_rate ($t-1$)  & 44.72 & 7.18 & 46.31 & 30.35 & 58.87 \\
unempl\_rate ($t-1$)  & 9.58 & 5.93 & 7.86 & 1.85 & 28.01 \\
obe\_p ($t-1$)  & 9.69 & 1.77 & 9.61 & 5.07 & 14.99 \\
obe\_65ov\_p ($t-1$)  & 13.69 & 2.99 & 13.45 & 5.38 & 22.82 \\
ovweight\_p ($t-1$)  & 34.63 & 3.13 & 34.48 & 26.55 & 41.80 \\
ovweight\_65ov\_p ($t-1$)  & 43.71 & 3.90 & 43.79 & 26.66 & 57.30 \\
smok\_p ($t-1$)  & 21.67 & 2.72 & 21.53 & 14.19 & 30.68 \\
smok\_65ov\_p ($t-1$)  & 9.46 & 1.99 & 9.39 & 4.24 & 15.68 \\
big\_smok\_p ($t-1$)  & 7.03 & 2.95 & 6.81 & 0.00 & 16.27 \\
gdp\_cap ($t-1$)  & 25,582.45 & 7,711.67 & 25,435.94 & 10,526.90 & 48,551.40 \\
co2\_cap ($t-1$)  & 8.91 & 2.79 & 8.30 & 3.15 & 17.31 \\
h\_beds ($t-1$)  & 3.97 & 0.98 & 3.71 & 2.12 & 8.19 \\
physicians ($t-1$)  & 2.01 & 0.28 & 2.00 & 1.14 & 3.14 \\
nurses ($t-1$)  & 4.63 & 0.69 & 4.70 & 3.05 & 6.68 \\
CT\_scan ($t-1$)  & 0.0266 & 0.0102 & 0.0257 & 0.0030 & 0.0575 \\
MRI\_scan ($t-1$)  & 0.0181 & 0.0114 & 0.0175 & 0.00 & 0.0554 \\
\midrule
he\_pri  & 504.88 & 151.17 & 492.00 & 214.00 & 1,054.00 \\
he\_pub  & 1,677.67 & 352.25 & 1,769.70 & 779.27 & 2,580.00 \\
he\_tot  & 2,182.54 & 464.68 & 2,235.00 & 1,018.27 & 3,359.00 \\

\end{tabular*}

\end{table}

\begin{table}[H]
    \centering
    \caption{Time windows splittings}
    \label{tab:timewindows}
    \begin{tabular}{lll}
        \hline
        \textbf{Time Window} & \textbf{Training set} & \textbf{Test set}\\
        \hline
        $time$-$wise$ & 1997-2013 & 2014-2019 \\
        $W_1$ & 1997-2005 & 2006-2007 \\
        $W_2$ & 2000-2007 & 2008-2009 \\
        $W_3$ & 2001-2009 & 2010-2011 \\
        $W_4$ & 2010-2017 & 2018-2019 \\
    \end{tabular}
\end{table}

\begin{table}[h]
    \centering
        \caption{Unique occurrences of the top-5 most frequent variables for $he\_pri$}
    \label{tab:freqhepri}
    \begin{tabular}{|c|c|c|c|c|c|}
      \hline
      \textbf{Variable} & \textbf{time-wise} & \textbf{$W_1$} & \textbf{$W_2$} & \textbf{$W_3$} & \textbf{$W_4$}\\ \hline
      $he\_pub\_p_{t-1}$ & \cellcolor{green!72!black}95 & - & \cellcolor{green!65!black}83 & \cellcolor{green!75!black}100 & \cellcolor{green!75!black}100\\ \hline
      $he\_pub\_p_{t-2}$ & - & \cellcolor{green!72!black}95 & - & - & -\\ \hline
      $a\_hh\_size_{t-2}$ & \cellcolor{green!66!black}85 & \cellcolor{green!43!white}47 & - & \cellcolor{green!71!black}94 & \cellcolor{green!42!white}45\\ \hline
      $a\_hh\_size_{t-1}$ & - & - & \cellcolor{green!69!black}90 & - & -\\ \hline
      $CT\_scan_{t-2}$ & - & - & \cellcolor{green!68!black}89 & \cellcolor{green!44!white}49 & -\\ \hline
      $MRI\_scan_{t-2}$ & \cellcolor{green!63!black}80 & \cellcolor{green!45!white}50 & - & \cellcolor{green!42!white}45 & -\\ \hline
      $str\_r\_p_{t-1}$ & \cellcolor{green!58!black}71 & - & - & - & -\\ \hline
            $str\_r\_p_{t-2}$ & - & \cellcolor{green!46!white}51 & - & - & \cellcolor{green!44!white}49\\ \hline
      $act\_rate_{t-2}$ & - & \cellcolor{green!56!black}69 & \cellcolor{green!49!white}56 & - & -\\ \hline

      $gdp\_cap_{t-2}$ & \cellcolor{green!29!white}23 & - & - & \cellcolor{green!42!white}45 & \cellcolor{green!36!white}35\\ \hline
      $nurses_{t-2}$ & - & - & \cellcolor{green!37!white}36 & - & -\\ \hline
      $empl\_rate_{t-2}$ & - & - & - & - & \cellcolor{green!33!white}30\\ \hline
    \end{tabular}

\end{table}

{
\scriptsize
\renewcommand{\arraystretch}{1.15}
\setlength{\tabcolsep}{3pt}
\begin{table}[H]
\centering
\begin{tabularx}{\linewidth}{|
>{\centering\arraybackslash}p{1.7cm}|
>{\raggedright\arraybackslash}X|
>{\centering\arraybackslash}p{0.8cm}|
>{\centering\arraybackslash}p{1.0cm}|}
\hline
\textbf{Split Type} & \textbf{Expression} & $R^2$ & \textit{MAPE} \\ \hline

\multirow{3}{*}{Time-wise} &
$\begin{aligned}[t]
he\_pri &= 0.41*gdp\_cap_{t-2} - 0.54*he\_pub\_p_{t-1} \\
        &\quad - 0.25*h\_beds_{t-2} + 0.38*MRI\_scan_{t-2}
\end{aligned}$
& 0.82 & 7.21 \\

&
$\begin{aligned}[t]
he\_pri &= -0.15*a\_hh\_size_{t-2} + 0.33*gdp\_cap_{t-2} \\
        &\quad - 0.50*he\_pub\_p_{t-1} - 0.23*h\_beds_{t-2} \\
        &\quad + 0.35*MRI\_scan_{t-2}
\end{aligned}$
& 0.81 & 7.37 \\

&
$\begin{aligned}[t]
he\_pri &= -0.15*a\_hh\_size_{t-2} + 0.33*gdp\_cap_{t-2} \\
        &\quad - 0.50*he\_pub\_p_{t-1} - 0.23*h\_beds_{t-2} \\
        &\quad + 0.35*MRI\_scan_{t-2}
\end{aligned}$
& 0.81 & 7.39 \\ \hline

\multirow{3}{*}{$W_1$} &
$\begin{aligned}[t]
he\_pri &= -1.1*gdp\_cap_{t-1} + 1.58*gdp\_cap_{t-2} \\
        &\quad - 0.44*he\_pub\_p_{t-2} + 0.23*MRI\_scan_{t-1} \\
        &\quad + 0.21*p\_r\_75ov\_p_{t-2}
\end{aligned}$
& 0.83 & 6.65 \\

&
$\begin{aligned}[t]
he\_pri &= 0.54*gdp\_cap_{t-2} - 0.44*he\_pub\_p_{t-2} \\
        &\quad - 0.16*h\_beds_{t-2} + 0.20*MRI\_scan_{t-1} \\
        &\quad - 0.16*p\_r\_m\_p_{t-2}
\end{aligned}$
& 0.82 & 6.32 \\

&
$\begin{aligned}[t]
he\_pri &= 0.54*gdp\_cap_{t-2} - 0.42*he\_pub\_p_{t-2} \\
        &\quad + 0.12*CT\_scan_{t-1} + 0.13*MRI\_scan_{t-1} \\
        &\quad + 0.20*p\_r\_75ov\_p_{t-2}
\end{aligned}$
& 0.82 & 6.65 \\ \hline

\multirow{3}{*}{$W_2$} &
$\begin{aligned}[t]
he\_pri &= -0.20*a\_hh\_size_{t-2} + 0.35*gdp\_cap_{t-2} \\
        &\quad - 0.51*he\_pub\_p_{t-1} \\
        &\quad + 0.93*\sqrt{0.61*MRI\_scan_{t-1} + 1.2} - 0.98
\end{aligned}$
& 0.80 & 6.06 \\

&
$\begin{aligned}[t]
he\_pri &= -0.18*a\_hh\_size_{t-1} + 0.29*gdp\_cap_{t-2} \\
        &\quad - 0.47*he\_pub\_p_{t-1} + 0.15*nurses_{t-1} \\
        &\quad + 0.23*MRI\_scan_{t-1}
\end{aligned}$
& 0.80 & 6.42 \\

&
$\begin{aligned}[t]
he\_pri &= -0.18*a\_hh\_size_{t-1} + 0.29*gdp\_cap_{t-2} \\
        &\quad - 0.47*he\_pub\_p_{t-1} + 0.14*nurses_{t-1} \\
        &\quad + 0.23*MRI\_scan_{t-1}
\end{aligned}$
& 0.80 & 6.42 \\ \hline

\multirow{3}{*}{$W_4$} &
$\begin{aligned}[t]
he\_pri &= 0.15*he\_pub\_p_{t-1}^{2} \\
        &\quad + 0.17*he\_pub\_p_{t-1}*nurses_{t-2} \\
        &\quad - 0.75*he\_pub\_p_{t-1} + 0.20*nurses_{t-2} - 0.09
\end{aligned}$
& 0.95 & 3.92 \\

&
$\begin{aligned}[t]
he\_pri &= 0.26*gdp\_cap_{t-2} - 0.86*he\_pub\_p_{t-1} \\
        &\quad + 0.34*empl\_rate_{t-2} - 0.48*str\_r\_p_{t-2} \\
        &\quad + 0.15*mort\_rate_{t-2}
\end{aligned}$
& 0.94 & 3.89 \\

&
$\begin{aligned}[t]
he\_pri &= 0.82*(0.17*he\_pub\_p_{t-1} - 1.1) \\
        &\quad *(-0.34*gdp\_cap_{t-2} + 0.76*he\_pub\_p_{t-1} \\
        &\quad + 0.18*pop\_res_{t-1}) - 0.13
\end{aligned}$
& 0.94 & 3.92 \\ \hline

\end{tabularx}
\caption{A sample of high-performing models for $he\_pri$ for each splitting strategy, based on $R^2$ and \textit{MAPE} scores.}
\label{tab:best-models}
\end{table}
}

\begin{table}[H]
\centering

\begin{subtable}[]{0.90\textwidth}
\centering
\begin{tabular}{lcccc}
\toprule
 & \multicolumn{2}{c}{$R^2$} & \multicolumn{2}{c}{MAPE (\%)}\\
\cmidrule(lr){2-3} \cmidrule(lr){4-5}
Type & Test & Out-of-time Evaluation & Test & Out-of-time Evaluation\\
\midrule
time-wise & 0.76 & 0.51 & 8.49 & 10.01\\
$W_1$ & 0.73 & -7.51 & 8.41 & 14.97 \\
$W_2$ & 0.74 & 0.15  & 7.29 & 13.35\\
$W_3$ & 0.25 & 0.22 & 15.57 & 12.34\\
$W_4$ & 0.92 & \textbf{0.76} & 4.25 & \textbf{7.12}\\
\bottomrule
\end{tabular}
\caption{Out-of-time evaluation of $he\_pri$ models on $2020$ year}
\label{tab:eval-pri-2020}
\end{subtable}

\begin{subtable}[H]{0.90\textwidth}
\centering
\begin{tabular}{lcccc}
\toprule
 & \multicolumn{2}{c}{$R^2$} & \multicolumn{2}{c}{MAPE (\%)}\\
\cmidrule(lr){2-3} \cmidrule(lr){4-5}
Type & Test & Out-of-time Evaluation & Test & Out-of-time Evaluation\\
\midrule
time-wise & 0.76 & -0.03 & 8.49 & 15.88\\
$W_1$ & 0.73 & -1.27 & 8.41 & 17.77 \\
$W_2$ & 0.74 & -0.86  & 7.29 & 25.66\\
$W_3$ & 0.25 & -0.81 & 15.57 & 25.31\\
$W_4$ & 0.92 & 0.16 & 4.25 & 18.23\\
\bottomrule
\end{tabular}
\caption{Out-of-time evaluation of $he\_pri$ models on $2021$ year}
\label{tab:eval-pri-2021}
\end{subtable}

\end{table}

\begin{table}[H]
\centering
\renewcommand{\arraystretch}{1.8}
\begin{tabularx}{\textwidth}{|X|c|c|}
\hline
 \textbf{Expression}  & $R^2$ & \textit{MAPE} \\ \hline
\footnotesize $ he\_pri =-0.19*a\_hh\_size_{t-1} - 0.81*he\_pub\_p_{t-1}  + 0.46*empl\_rate_{t-2} - 0.44*str\_r\_p_{t-2}$         & 0.90  & 5.85   \\ \hline
\footnotesize $ he\_pri =-0.19*a\_hh\_size_{t-1} - 0.81*he\_pub\_p_{t-1}  + 0.46*empl\_rate_{t-2} - 0.44*str\_r\_p_{t-2}$       & 0.90  & 5.85   \\ \hline
\footnotesize $ he\_pri =-0.89*he\_pub\_p_{t-1} + 0.52*empl\_rate_{t-2}+ 0.62*str\_r\_p_{t-2}/(0.53*empl\_rate_{t-2} - 1.4) + 0.128 $        & 0.90  & 6.09  \\ \hline
\footnotesize $ he\_pri = -0.07*area_{t-2} - 0.18*a\_hh\_size_{t-2} - 0.81*he\_pub\_p_{t-1}  + 0.43*empl\_rate_{t-2} - 0.41*str\_r\_p_{t-2}$        & 0.90  & 5.46    \\ \hline
                           \footnotesize $ he\_pri = -0.07*area_{t-2} - 0.18*a\_hh\_size_{t-2} - 0.80*he\_pub\_p_{t-1}  + 0.43*empl\_rate_{t-2} - 0.41*str\_r\_p_{t-2}$                     & 0.90  & 5.48        \\ \hline
                           \footnotesize  $ he\_pri = -0.07*area_{t-2} - 0.18*a\_hh\_size_{t-2} - 0.80*he\_pub\_p_{t-1}  + 0.43*empl\_rate_{t-2} - 0.41*str\_r\_p_{t-2}$                               & 0.90  & 5.48     \\ \hline
                          \footnotesize  $ he\_pri = -0.07*area_{t-2} - 0.19*a\_hh\_size_{t-2} - 0.80*he\_pub\_p_{t-1} + 0.42*empl\_rate_{t-2} - 0.40*str\_r\_p_{t-2}$                              & 0.90  & 5.50   \\ \hline
                        \footnotesize  $ he\_pri = -0.20*a\_hh\_size_{t-2} - 0.81*he\_pub\_p_{t-1} + 0.45*empl\_rate_{t-2} - 0.44*str\_r\_p_{t-2}$                         & 0.89  & 5.86  \\ \hline
                          \footnotesize  $ he\_pri = -0.20*a\_hh\_size_{t-2}  - 0.81*he\_pub\_p_{t-1} + 0.45*empl\_rate_{t-2} - 0.44*str\_r\_p_{t-2}$                               & 0.89  & 5.86  \\ \hline
                        \footnotesize  $ he\_pri = -0.20*a\_hh\_size_{t-2}  - 0.81*he\_pub\_p_{t-1} + 0.45*empl\_rate_{t-2} - 0.44*str\_r\_p_{t-2}$                               & 0.89  & 5.86  \\ \hline
\end{tabularx}
\caption{Top-10 $he\_pri$ models over $2020$}
\label{tab:best-models-eval}
\end{table}

\begin{table}[]
\centering
\renewcommand{\arraystretch}{1.8}
\begin{tabularx}{\textwidth}{|X|c|c|c|}
\hline
 \textbf{Substructure}  & Frequency ($\%$) & $R^2$ & \textit{MAPE} \\ \hline
 $C*he\_pub\_p_{t-1} +  C*str\_r\_p_{t-2}$    &  46  & 0.93  & 4.24   \\ \hline
$C*a\_hh\_size_{t-2} + C*he\_pub\_p_{t-1}$    &  41  & 0.93  & 4.21  \\ \hline
$C*gdp\_cap_{t-2} + C*he\_pub\_p_{t-1}$    &  30  & 0.94  & 3.98  \\ \hline
$ C*he\_pub\_p_{t-1} + C*empl\_rate_{t-2}$    &  29  & 0.92  & 4.21  \\ \hline
$C*co2\_cap_{t-1} + C*he\_pub\_p_{t-1}$    &  25  & 0.93  & 4.19  \\ \hline
$C*a\_hh\_size_{t-2} + C*str\_r\_p_{t-2}$    &  24  & 0.94  & 3.97 \\ \hline
$C*a\_hh\_size_{t-2} + C*he\_pub\_p_{t-1} + C*str\_r\_p_{t-2}$    &  24  & 0.94  & 3.97 \\ \hline
$ C*a\_hh\_size_{t-2} + C*empl\_rate_{t-2}$    &  21 & 0.93  & 3.99  \\ \hline
$ C*empl\_rate_{t-2} + C*str\_r\_p_{t-2}$    &  20 & 0.94  & 3.90  \\ \hline
$ C*he\_pub\_p_{t-1} + C*empl\_rate_{t-2} + C*str\_r\_p_{t-2}$    &  20  & 0.94  & 3.90 \\ \hline
$C*a\_hh\_size_{t-2}+ C*he\_pub\_p_{t-1}  +  C*empl\_rate_{t-2}$    &  20  & 0.93  & 3.97 \\ \hline
$ C*a\_hh\_size_{t-2} +  C*empl\_rate_{t-2} + C*str\_r\_p_{t-2}$    &  17  & 0.94  & 3.89   \\ \hline
$ C*a\_hh\_size_{t-2} + C*he\_pub\_p_{t-1} + C*empl\_rate_{t-2} + C*str\_r\_p_{t-2}$    &  17  & 0.94  & 3.89   \\ \hline
$ C*co2\_cap_{t-1} + C*gdp\_cap_{t-2} + C*he\_pub\_p_{t-1} $    &  16  & 0.94  & 3.84   \\ \hline

\end{tabularx}
\caption{Top $15$ frequent substructures for expressions obtained in $W_4$ time window. $R^2$ and \textit{MAPE} values correspond to the average error scores obtained when that substructure is present.}
\label{tab:freq-substructures}
\end{table}

\pagebreak

\section*{Statements and Declarations}

\subsection*{Funding}
No funds, grants, or other support were received for conducting this study or preparing this manuscript.

\subsection*{Competing interests}
The authors have no relevant financial or non-financial interests to disclose.

\subsection*{Ethics approval and consent to participate}
This study used aggregated regional administrative data and did not involve human participants, human material, or animals. Ethics approval and informed consent to participate were therefore not required.

\subsection*{Consent for publication}
Not applicable.

\subsection*{Data availability}
The processed data supporting the findings of this study are available from the corresponding author upon reasonable request. The underlying source data were obtained from the Italian National Institute of Statistics (Istat) Health for All database and Istat statistics, subject to the applicable terms of use.

\subsection*{Materials availability}
Materials supporting the findings of this study are available from the corresponding author upon reasonable request.

\subsection*{Code availability}
The code used to conduct data processing and the reported analyses is available from the corresponding author upon reasonable request.

\subsection*{Author contributions}
All the authors conceived the study. Cristina Rossetti and Nicola Caravaggio developed the theoretical framework and drafted the manuscript. Nicola Caravaggio performed data collection and curation. Cristina Rossetti performed data cleaning and carried out the experiments. Giovanni Lamura and Giuliano Resce supervised the project and critically revised the final version of the manuscript. All authors discussed the results, contributed to the final manuscript, and approved the submitted version.

\end{document}